\documentclass[12pt]{iopart}

\usepackage[colorlinks=true,urlcolor=blue,citecolor=blue,linkcolor=blue]{hyperref}
\usepackage{graphicx}
\expandafter\let\csname equation*\endcsname\relax
\expandafter\let\csname endequation*\endcsname\relax
\usepackage{amsmath}
\usepackage{amsfonts}
\usepackage{cite} 

\begin{document}

\title[Spectroscopy and critical quantum thermometry in the ultrastrong coupling regime]
{Spectroscopy and critical quantum thermometry in the ultrastrong coupling regime}

\author{M.  Salado-Mej\'ia, R. Rom\'an-Ancheyta$^*$,  F. Soto-Eguibar and H. M. Moya-Cessa }

\address{Instituto Nacional de Astrof\'isica, \'Optica y Electr\'onica, Calle Luis Enrique Erro 1, Santa Mar\'ia
Tonantzintla, Puebla, 72840, Mexico}

\ead{$^*$ancheyta6@gmail.com}


\begin{abstract}
We present an exact analytical solution of the anisotropic Hopfield model, and we use it to investigate in detail the spectral and thermometric response of two ultrastrongly coupled quantum systems. Interestingly, we show that depending on the initial state of the coupled system, the vacuum Rabi splitting manifests significant asymmetries that may be considered spectral signatures of the counterintuitive decoupling effect. Using the coupled system as a thermometer for quantum thermodynamics applications, we obtain the ultimate bounds on the estimation of temperature that remain valid in the ultrastrong coupling regime. Remarkably, if the system performs a quantum phase transition, the quantum Fisher information exhibits periodic divergences, suggesting that one can have several points of arbitrarily high thermometric precision for such a critical quantum sensor.
\end{abstract}

\section{Introduction}
When the coupling strength between light and matter starts to be comparable with the system's natural frequencies, the ultrastrong coupling (USC) regime of the light-matter interaction emerges~\cite{FriskKockum2019}. 
The USC regime has received considerable attention from the theoretical and experimental point of view \cite{Forn-Diaz2019} during the last decade, mainly because it enables more efficient interactions that one could potentially use in quantum technologies with hybrid systems~\cite{Kurizki3866}.

The USC regime is characterized,  among other things, for the breakdown of the rotating-wave approximation 
(RWA)~\cite{Boite2020}, an approximation commonly used in quantum optics. Unambiguous evidence of this breakdown has been acquired 
through  spectroscopy measurements in several experiments, see for instance Refs.~\cite{Niemczyk2010,Li2018NatPhot,Makihara_arXiv_20}. From a theoretical point of view, there are different ways to
determine the spectral response of such ultrastrongly coupled quantum systems; for example, one might calculate the transmission or absorption spectra using the input-output theory~\cite{Ciuti2006} or through the Wiener–Khintchine power spectrum using the theory of open quantum systems~\cite{breuer2002theory}.
A detailed treatment of the interaction between the 
coupled system and its environment must be done in 
either case. 
For the latter, a full microscopic derivation, sometimes 
called a global approach, of the corresponding master equation  
is the standard procedure to follow~\cite{Hofer_2017}.
However, such derivation is highly dependent on either 
the coupled system is in the strong or ultrastrong 
coupling regime~\cite{Blais2011}, and if the Born, 
Markov, or secular approximations have to be considered.
Those approaches may then force one to opt for 
numerical solutions that sometimes make difficult 
to obtain valuable information about the problem. 
Here, we advocate on using the so-called 
time-dependent physical spectrum of light introduced
by Eberly and W\'odkiewicz (EW) in~\cite{Eberly1977}, 
which is based on, somehow, 
\textcolor{black}{an operational} approach. Remarkably, we find that 
for the case where there is a single excitation in the 
coupled system, which is a common situation for several 
up to date experiments, the EW spectrum can describe, in 
simple terms, some of the most interesting effects predicted 
to happen in the USC regime of the light-matter interaction.

On the other hand, it is known that spectroscopy measurements 
of a quantum system can also be used to extract information 
about the temperature of the thermal environment the system 
is in contact with. This strategy has been experimentally 
implemented with the name of fluorescence 
thermometry~\cite{Haupt,Kuckso}.
Recently, such idea was extended to the notion of the 
so-called {apparent temperature} associated with non-thermal 
baths~\cite{roman2019spectral,Latune_2019}, i.e., baths having 
a certain amount of quantum coherence.
Along these lines of research, in this work, we use the 
Hopfield model~\cite{Hopfield1958} as an example of a 
composite quantum probe 
acting as a thermometer~\cite{Deffner2019book,Correa_Review}, 
and we obtain fundamental bounds on the precision of temperature 
estimation. In contrast to previous works where the RWA is 
used~\cite{Correa_Review,Campbell2017},
our quantum thermometry results remain valid in the USC regime. 
Moreover, we exploit the thermometer's criticality when, close 
to the critical point of a quantum phase transition, the 
corresponding quantum Fisher information, which is proportional 
to the square of the signal-to-noise ratio, diverges.

The organization of the paper is the following: in Sec.~\ref{HM},
we introduce the anisotropic Hopfield model and present a general 
analytical solution of the corresponding eigenfrequencies,
polaritons, that we use to describe the energy level structure 
of the coupled system. 
In Sec.~\ref{SHM}, we compute the corresponding EW spectrum and get 
a vacuum Rabi splitting with large asymmetries that can be considered 
spectral signatures of the counterintuitive decoupling effect.
Sec.~\ref{QT} deals with the quantum thermometry analysis, where 
we use the Hopfield model as a thermometer for the estimation of 
temperature. Sec.~\ref{Conc} shows the conclusions.


\section{The Hopfield model}\label{HM}

We start by writing the Hamiltonian of the simplified version 
of the Hopfield model~ \cite{FriskKockum2019,Boite2020}:
\begin{equation}\label{hopfield}
\hat{H}_{\rm Hopfield}=\omega_c \hat{a}^\dagger \hat{a} + \omega_b \hat{b}^\dagger \hat{b} + i g_1 (\hat{a}\hat{b}^\dagger - \hat{a}^\dagger \hat{b} ) 
+ i g_2 (\hat{a}^\dagger\hat{b}^\dagger - \hat{a} \hat{b} ) 
+ D ( \hat{a}+\hat{a}^\dagger)^2,
\end{equation}
where we have set $\hbar=1$.
This kind of Hamiltonian is widely used to describe, 
in the weak excitation limit, several experiments 
dealing with the USC between two effective quantum
systems~\cite{Mueller2020,Li2018NatPhot,Makihara_arXiv_20,Zhang2016,Keller2020,ScalariScience,DeLiberato2010,Bamba2019,Maissen_2017}. 
For instance, in the light-matter interaction,
the coupling could be between a highly confined single electromagnetic 
field mode of frequency $\omega_c$, with a collective excitation of 
a matter system of frequency 
$\omega_b$, which are described by the free (unperturbed) Hamiltonians 
$\omega_c\hat{a}^\dagger\hat{a}$ and $\omega_b\hat b^\dagger\hat b$
respectively. Often, the matter system 
is an ensemble made of a large number of two-level systems 
that can be bosonize~\cite{Bamba2019}. In such case, 
$\hat{b}$ ($\hat{a}$) and $\hat{b}^\dagger$($\hat{a}^\dagger$) 
will represent, respectively, 
the annihilation and creation matter (field) operators with
the usual commutation relation, 
$[\hat{b},\hat{b}^\dagger]=[\hat{a},\hat{a}^\dagger]=1$.
On the third and four-term of Eq.~(\ref{hopfield}), $g_1$ 
and $g_2$ are the coupling strengths of the 
commonly named corotating and counterrotating terms, respectively%
\textcolor{black}{. Following historical convention~\cite{FriskKockum2019,Forn-Diaz2019}, we can define the USC regime of the anisotropic Hopfield model when $0.1{\omega_c}\leq {g}_{1,2}\leq {\omega_c}$~\cite{Makihara_arXiv_20}, where we have assumed resonant conditions $\omega_c=\omega_b$. The counterrotating terms in $\hat{H}_{\rm Hopfield}$}
should not be neglected anymore if the 
coupled system is in the USC regime. 
Recall that when light interacts with natural atoms, $g_1$ 
must be equal to $g_2$. 
However, recent experiments on matter-matter interactions
(specifically magnon-magnon interaction) show that one 
can have \textcolor{black}{surprising} situations \textcolor{black}{where} $g_2>g_1$~\cite{Makihara_arXiv_20,Makihara20}. 
There is also evidence that such anisotropy ($g_1\neq g_2$) 
can better describe the experimental data of the USC between 
electromagnetic radiation with artificial atoms made of 
superconducting quantum circuits~\cite{Xie2014}.
Thus, $\hat{H}_{\rm Hopfield}$ in Eq.~(\ref{hopfield}) 
might be called the anisotropic Hopfield model in analogy 
with the anisotropic Rabi model~\cite{Xie2014,Romero_2018}.
The last term in Eq.~(\ref{hopfield}) is the so-called 
diamagnetic term (because it is responsible for diamagnetism) 
associated with the square of the vector potential of the 
field and can be viewed as a self-interaction energy with 
$D$ its strength~\cite{FriskKockum2019,Forn-Diaz2019}. 
This term is relevant and can be even dominant in the 
USC regime of the light-matter interaction. However, in some 
matter-matter interaction experiments, like the 
spin-magnon~\cite{Li2018} or magnon-magnon~\cite{Makihara_arXiv_20}, 
such term may not appear. 
Interestingly, although the anisotropic Hopfield 
model does not conserve the total number of excitations 
\cite{FriskKockum2019}, it still preserves a discrete 
symmetry, the $\mathbb{Z}_2$ or parity symmetry~\cite{Braak_2019}, 
because $\hat{H}_{\rm Hopfield}$ commutes with the 
parity operator 
$\hat{P}=\exp\big[i\pi(\hat{a}^\dagger\hat{a}+\hat{b}^\dagger\hat{b})\big]$,
i.e., the Hamiltonian is invariant under the transformation 
$\hat{P}^\dagger\hat{H}_{\rm Hopfield}\hat{P}=\hat{H}_{\rm Hopfield}$,
which means that spectral crossings between energy levels
from different symmetry sectors are allowed to 
occur~\cite{li2020hidden}.

It is well known that the energy level structure of
$\hat{H}_{\rm Hopfield}$ is quite rich, and in the USC 
regime, it has substantial deviations from the standard 
energy spectrum obtained when the RWA is considered. 
This fact has been shown numerically~\cite{FriskKockum2019} 
and analytically%
~\cite{Tufarelli2015} for $g_1=g_2$ and
$D\neq 0$. 
Attempts to generalize the previous results for $g_1\neq g_2$ were made very recently in~\cite{Ivette2019}
where, through a relatively complicated procedure, an analytical result 
of the eigenfrequencies was obtained. 
In the following, we show  that, indeed, it is possible to 
get a complete analytical solution of the eigenvalues of 
$\hat{H}_{\rm Hopfield}$ in the most general case, i.e.,
when $g_1\neq g_2$, $D\neq 0$ and the coupled system is 
off-resonance $\omega_c\neq\omega_b$.

Since $\hat{H}_{\rm Hopfield}$ 
is a quadratic Hamiltonian, we can use a 
series of simple, but by no means trivial, unitary transformations 
to diagonalize it. We only need a combination of two 
rotations and one squeezing transformation that allow us to write the eigenmodes
of $\hat{H}_{\rm Hopfield}$ as the Hamiltonian of two uncoupled 
harmonic oscillators with frequencies 
(see \ref{app1} for a detailed derivation):
\begin{equation}\label{polfre}
2\,\omega_{x,y}^2=(2 \lambda _1\lambda_2+\Omega _x^2+\Omega _y^2)\pm \sqrt{\left(1-\lambda _2^2\right) (\Omega _x^2-\Omega _y^2){}^2+[2 \lambda _1+\lambda _2 (\Omega _x^2+\Omega _y^2)]^2},
\end{equation}
where 
$\Omega_x^2=\omega_c^2+4 D \omega_c$, $\Omega_y^2$=$\omega_b^2$ and 
$\lambda_{1,2}=\left( g_1 \pm g_2 \right)({\omega_c \omega_b})^{\pm {1}/{2}}$.
Therefore, the exact eigenvalues of $\hat{H}_{\rm Hopfield}$
are $E_{mn}=\omega_x(m+1/2)+\omega_y(n+1/2)$, with $n$ and $m$ 
non-negative integers. In the context of light-matter interaction,
$\omega_{x,y}$ are known as the upper and lower 
frequencies of the photonic quasiparticles called polaritons~\cite{Kaminer2020light}, hybrid light-matter states. 
As expected, Eq.~(\ref{polfre}) contains, as a particular case,
the results of~\cite{Tufarelli2015,Zhou2020} and~\cite{Ivette2019} when $g_1=g_2$ or $D=0$.

The analytical expressions of $\omega_{x,y}$ in Eq.~(\ref{polfre}) 
are useful for finding out the exact critical values of the 
parameters in $\hat{H}_{\rm Hopfield}$ that suppress or permit a 
possible quantum phase transition in the model. For instance, 
in order to suppress it, it is easy to show that the diamagnetic 
term should satisfy the inequality 
$D> D_{\rm crit}\equiv{(g_1+g_2){}^2}/({4 \omega _b})-{\omega _c}/{4}$,
implying that $\hat{H}_{\rm Hopfield}$ is bounded from below.
For the particular case where $g_1=g_2\equiv g$, $D_{\rm crit}$ 
reduces to the one obtained in~\cite{Rossi2017}. Additionally, 
if $D=0$, $\hat{H}_{\rm Hopfield}$ 
\textcolor{black}{resembles the effective Dicke Hamiltonian in 
the normal phase and in the thermodynamic limit with a quantum 
phase transition at
$g=g_{\rm crit}\equiv\sqrt{\omega_c\omega_b}/2$.
However, in the superradiant phase the corresponding effective
Hamiltonian of the Dicke model substantially differs from 
$\hat{H}_{\rm Hopfield}$, see~\cite{EmaryPR2003} and references
therein for more details.}

\begin{figure}[b]
\begin{centering}
\includegraphics[width=13cm, height=9cm]{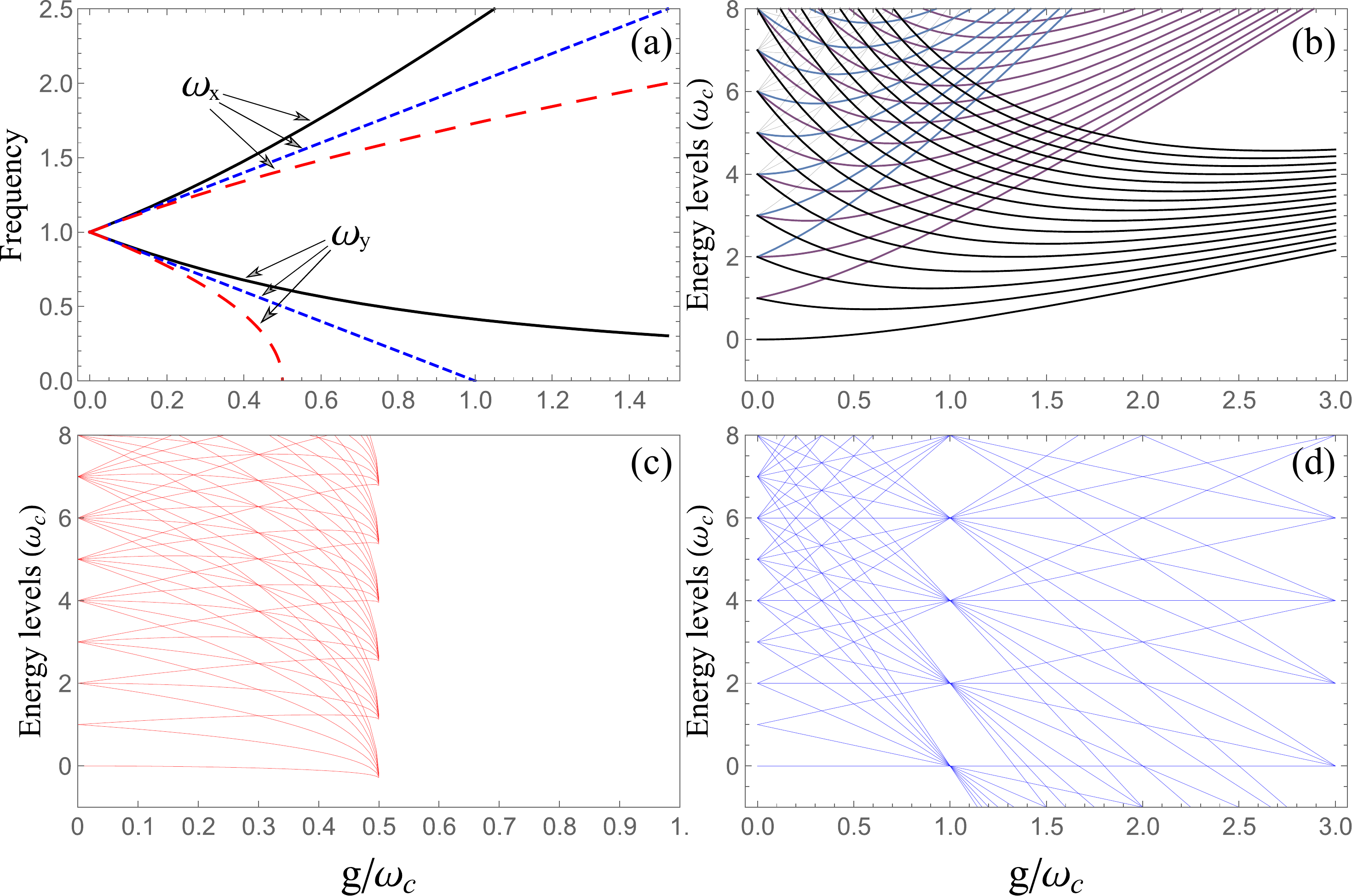}
\caption{$\bf{(a)}$ Polaritonic frequencies $\omega_{x,y}$ of $\hat{H}_{\rm Hopfield}$, cf. Eq.~\eqref{polfre} as a function of the normalized coupling strength $g/\omega_c$. Under resonance conditions ($\omega_c=\omega_b$), we set $g_1=g_2= g$ with $D=g^2/\omega_b$ (black solid lines) and $D=0$ (red dashed lines). Blue dotted lines are the frequencies obtained 
after the RWA is applied in Eq.~(\ref{polfre}). $\bf{(b)}$ Energy levels, $E_{mn}=\omega_x m+\omega_y n$ with $m,n\in \mathbb{Z}^+$, of $\hat{H}_{\rm Hopfield}$ versus $g/\omega_c$. The corresponding set of parameters are the same as those of the black solid lines in $\bf{(a)}$. $\bf{(c)}$ Same as $\bf{(b)}$ but now the parameters are those of the red dashed lines in $\bf{(a)}$. An evident spectral collapse occurs because $\omega_y$ is pushed to zero at $g_{\rm crit}=\omega_c/2$ where the system exhibits a super radiant phase transition~\cite{FriskKockum2019}. $\bf{(d)}$ Same as $\bf{(b)}$ but using the parameters of the blue dotted lines of $\bf{(a)}$. As expected, all the energy-level diagrams coincide in the region $g/\omega_c\leq 0.1$, i.e.,  far away from the USC regime, see the main text for more details.}
\label{tutti}
\end{centering}
\end{figure}

It is instructive to see how the polaritonic frequencies 
$\omega_{x,y}$ and the energy eigenvalues $E_{mn}$ change 
when the coupled system starts to enter the USC regime. 
For instance, if $g_1=g_2=g$ 
(for simplicity), Eq.~(\ref{polfre}) reduces to
\begin{equation}\label{omega_dia}
2\,\omega_{x,y}^2=(\omega _c^2+\omega _b^2+4 D \omega _c) \pm \big[{(\omega _c^2-\omega _b^2 +4 D \omega _c){}^2+16 g^2 \omega _c \omega _b}\big]^{\frac{1}{2}}.
\end{equation}
Additionally, by considering the Thomas–Reiche–Kuhn (TRK) sum 
rule~\cite{Ciuti2005,Wang1999,Nataf2010} in light-matter 
interaction, the diamagnetic term takes the value 
$D= g^2/\omega_b$~\cite{Tufarelli2015,Nataf2010}. Thus, under 
resonant conditions and together with the TRK sum rule, 
we show in Fig.~\ref{tutti}$(a)$ the behavior of 
Eq.~(\ref{omega_dia}) as a function of the normalized coupling 
$g/\omega_c$ (solid black lines). There we can see how the 
frequency of the upper (lower) polariton $\omega_x$ ($\omega_y$),
as its name implies, increases (decreases) as a function of 
$g/\omega_c$. If $D=0$, which can happen for some of the 
cases mentioned above, the behavior of Eq.~(\ref{omega_dia})
is shown by the red dashed lines of Fig.~\ref{tutti}$(a)$.
Contrary to the previous case, the eigenfrequency $\omega_y$
is now pushed to zero by increasing the normalized coupling.
Then the coupled system undergoes a superradiant phase 
transition (SPT) (see~\cite{Viehmann2011,Andolina_PRB_2019,Andolina_PRB_2020} and references 
therein for a detailed discussion on the applicability of 
generalized no-go theorems in different light-matter scenarios 
like circuit-QED). 
It is illustrative to compare these two previous 
examples with the one in which the RWA has been applied, 
i.e., when the counterrotating and the 
diamagnetic terms in Eq.~(\ref{polfre}) are neglected, that is, when
$g_2=D=0$ and $g_1\equiv g$. 
Under the RWA, the eigenfrequencies of Eq.~(\ref{polfre}) 
acquire a very simple form	
$2\omega_{x,y}^{\textsl{\tiny RWA}}\equiv({\omega_c+\omega_b})\pm\sqrt{({\omega_c-\omega_b})^2+4g^2}$
shown as the two blue dotted linear asymptotes of
Fig.~\ref{tutti}$(a)$. As expected, 
one can see that the RWA is a good approximation for the 
upper and lower polaritonic frequencies only when the 
normalized coupling is small enough, i.e., 
when the coupled system is far away from the USC regime.

In Fig.~\ref{tutti}$(b)$-$(d)$, we show the corresponding 
energy levels structure as a function of $g/\omega_c$,  
for the three cases previously discussed. For example, 
Fig.~\ref{tutti}$(b)$ displays (approximately) the first 
$50$ energy levels $E_{mn}$ associated with the polaritonic 
frequencies $\omega_{x,y}$ corresponding to the solid black
lines of~Fig.~\ref{tutti}$(a)$.
One can see that this figure differs substantially from the figure~2.f of the recent 
review~\cite{FriskKockum2019} dealing with USC between light
and matter.
It seems that in~\cite{FriskKockum2019} the energy levels 
structure was calculated by standard numerical methods and, 
due to the lack of a fully converged numerical solution during 
the diagonalization process, fictitious avoided crossing energy 
levels can be observed (in Fig.~2f) for values  $g/\omega_c\approx 1.5$. 
Even more, for larger values of $g/\omega_c$, 
around $3$, for example, Ref.~\cite{FriskKockum2019} predicts an 
anharmonic behavior in the energy spectrum. 
All this is in contrast with the exact analytical result of the
energy levels displayed in Fig.~\ref{tutti}$(b)$, where 
due to the model's parity symmetry, no avoided crossing energy 
levels are observed, and for large values 
of the normalized coupling strength, the energy levels are 
equispaced and not anharmonic as shown in~\cite{FriskKockum2019}. 
In fact, in the next section, we will see in detail that such 
equispaced behavior is a manifestation of the so-called decoupling 
effect~\cite{DeLiberato2014} occurring in the deep-strong coupling 
(DSC) regime. 
\textcolor{black}{In contrast to the USC, the DSC is defined for 
$g/\omega_{c}> 1$~\cite{FriskKockum2019,Forn-Diaz2019}}.
Figure~\ref{tutti}$(c)$ shows the energy levels $E_{mn}$ with 
frequencies $\omega_{x,y}$ associated with the red dashed lines 
of Fig.~\ref{tutti}$(a)$, i.e., when $D=0$ in Eq.~(\ref{omega_dia}).
There, one can see a spectral collapse~\cite{Duan2016} at the
critical coupling constant $g_{\rm crit}=\omega_c/2$; this happens 
because the corresponding $\omega_y$ is pushed to zero when the
normalized coupling increases and the coupled system enters the
SPT. Once again, in Fig. 2.f of Ref.~\cite{FriskKockum2019} 
it is difficult  to distinguish such spectral collapse due to an 
apparent deficient numerical solution.
Figure~\ref{tutti}$(d)$ also shows $E_{mn}$ but using
the eigenfrequencies $\omega_{x,y}^{\textsl{\tiny RWA}}$,
that correspond to the blue dotted lines of Fig.~\ref{tutti}$(a)$,
i.e., when the RWA is applied. Note that Figs.~\ref{tutti}$(b)$-$(d)$ are shifted by adding a factor 
$-({\omega_c+\omega_b})/{2}$ to $E_{mn}$ and, as expected, 
all of them coincide in the region 
$g/\omega_c\leq 0.1$~\cite{FriskKockum2019,Forn-Diaz2019}. 

\section{Spectroscopy of the Hopfield model} \label{SHM}


The Eberly-W\'odkiewicz (EW) time-dependent physical spectrum 
is defined as~\cite{Eberly1977, Eberly1980}:
\begin{equation}\label{spectrum}
S(\omega,t,\Gamma )=2 \Gamma  e^{-2 \Gamma t} \int_0^t dt_1 \int_0^t dt_2  e^{(\Gamma -i\omega)t_1} e^{(\Gamma +i\omega)t_2}\langle \hat{a}_1^{\dagger }(t_1)\hat{a}_1 (t_2) \rangle,
\end{equation}
where $\omega$ and $\Gamma$ are the central 
frequency and band half-width of a Fabry–P\'erot \textcolor{black}{interferometer} acting
as a filter and
$\langle \hat{a}_1^{\dagger }(t_1)\hat{a}_1 (t_2) \rangle$ 
is the field's autocorrelation function with the average 
done with respect to the initial state of the coupled system. 
If one is interested in the spectrum corresponding to the 
matter part of the light-matter interaction, then the previous 
correlation function should be replaced by
$\langle \hat{b}_1^{\dagger }(t_1)\hat{b}_1 (t_2) \rangle$.
Here, we will not describe in full detail how to take
into account the interaction between the ultrastrong coupled 
system with an environment; instead, we will see that the
corresponding unitary evolution generated by the Hopfield Hamiltonian, $\hat U(t)=\exp(-it\hat{H}_{\rm Hopfield})$, is enough for our purposes.
\textcolor{black}{
However, we would like to stress that the EW definition in Eq.~(\ref{spectrum}) is not totally based on such unitary time-evolution. Rather, the definition is based on the microscopic fully quantum theory of light detection by Glauber and, importantly, the insertion of a frequency-sensitive device (the filter) in front of the photodetector. The use of a finite filter bandwidth, necessary on any spectrally resolved observation of a light field, makes the EW spectrum more realistic than those spectra assuming infinite spectral resolution as the Wiener–Khintchine stationary power spectrum does. In fact, due to such an operational approach in deriving the EW spectrum, this can be used to study intrinsically nonstationary light fields. For instance, in a completely lossless situation, the first theoretical prediction for the existence of the vacuum Rabi splitting was made using the EW spectrum in~\cite{Mondragon_PRL_1983} by choosing a suitable initial state for the atom-cavity system; its impressive experimental demonstration followed this in~\cite{Kimble_PRL_1992}. 
Since then, several theoretical~\cite{Nazir_NP_2017} and experimental works~\cite{Laucht_PRL_2009,Laucht_PRB_2010} have considered the spectrometer’s finite spectral resolution in the same way that Eberly and W\'odkiewicz originally proposed for theoretical comparison with their spectroscopic measurements; see also~\cite{Golub_PRL_1987} for explicit time-dependent experimental spectra of resonance fluorescence.
}

We first consider the case where the initial state of the 
coupled system is $|1,0\rangle\equiv|1\rangle\otimes|0\rangle$, 
which means that the field initially contains one excitation 
(one photon) while the matter part is in its ground~(vacuum) state.
Additionally, and for the sake of simplicity, we assume that 
$g_1=g_2=g$.
In such situation, we can show that the field's autocorrelation 
function is given by  (see~\ref{app2} for details):
$\langle 1,0|\hat{a}^{\dagger }(t_1)\hat{a} (t_2)|1,0\rangle=f^*_1( t_1 )  f_1( t_2 )+2f^*_2( t_1 )  f_2( t_2 )+f^*_4( t_1 )  f_4( t_2 )$,
with the time-dependent functions
$f_j(t)$=$\sum_{k=1}^2\exp(\pm i\omega_x t)\mu_{jk}+\sum_{k=3}^4\exp(\pm i\omega_y t)\mu_{jk}$
and the time-independent coefficients $\mu_{jk}$ defined 
in~(\ref{mu}) of ~\ref{app2}. Notice that each $f_j(t)$ has 
a simple time-dependence given by the sum of the exponentials 
$\exp(\pm i\omega_{x,y}t)$, where the frequencies $\omega_{x,y}$ 
are those given in Eq.~(\ref{omega_dia}).
%
%
By substituting this result in Eq.~(\ref{spectrum}), one
can efficiently compute the two corresponding time-integrals, 
where the contribution of several of 
the resulting time-dependent terms, in the long time limit
$\Gamma t\gg 1$, are negligible due to their fast oscillations.
The long time limit is when the spectrum has been stabilized and there are no significant changes. However, it should
not be confused with the steady-state, which 
occurs when dissipation processes are explicitly taken into account;
we refer the reader to Refs.~\cite{roman2019spectral} and 
\cite{RicTime} where this observation has been tested by analytical and numerical 
methods in other examples of the light-matter interaction.
Therefore, in the long time limit and with the particular 
initial conditions above-mentioned, the EW spectrum of 
Eq.~(\ref{spectrum}) can be very well approximated by the 
time-independent expression
$S(\omega,\Gamma)\equiv S_x(\omega,\Gamma)+S_y(\omega,\Gamma)$,
where we have defined the following terms:
\begin{equation}\label{spectrum10}
S_{x,y}(\omega,\Gamma)= \frac{\frac{1}{2}\Gamma_{x,y}}{\Gamma ^2+(\omega-\omega_{x,y} )^2},
\quad
\Gamma_{x,y}=\frac{\Gamma  h_{x,y}}{4 \omega _c^2 \omega _{x,y}^2} \big[(\omega _c+\omega _{x,y})^4+2 (\omega _c^2-\omega _{x,y}^2)^2\big],
\end{equation}
$h_{x,y}=\big[{1}\pm {\Omega }(4 \lambda ^2+\Omega ^2)^{-\frac{1}{2}}\big]^2/4$, $\Omega ={\omega_c}^2+4 D\omega_c-{\omega_b}^2$ 
and $\lambda =2 g \sqrt{\omega_b \omega_c}$.
Thus, for the initial state $|1,0\rangle$, which contains 
only a single excitation in the coupled system, the EW 
spectrum of the Hopfield Hamiltonian is just the sum of 
two Lorentzian line shape functions $S_{x,y}(\omega,\Gamma)$.
This analytical spectrum is valid in the USC regimen for 
any value of the diamagnetic term $D$, and it is one of our 
major results. 
It is illustrative to compare this result with the corresponding 
spectrum that is obtained 
under the RWA and the same initial state $|1,0\rangle$; in such case, the calculations are more straightforward 
than the previous ones and the EW spectrum reduces to 
$S_{\textsl{\tiny RWA}}(\omega,\Gamma)= S_x^{\textsl{\tiny RWA}}(\omega,\Gamma)+S_y^{\textsl{\tiny RWA}}(\omega,\Gamma)$,
where
\begin{equation}\label{ho2}
S_{x,y}^{\textsl{\tiny RWA}}(\omega,\Gamma)=\frac{\frac{1}{2}\Gamma_{\textsl{\tiny RWA}}}{\Gamma^2+\big(\omega-\omega_{x,y}^{\textsl{\tiny RWA}}\big)^2},
\quad
\Gamma_{\textsl{\tiny RWA}}=\frac{4\Gamma g^2}{(\omega_b-\omega_c)^2+4g^2}\qquad\qquad\quad
\end{equation}
and 
$2\omega_{x,y}^{\textsl{\tiny RWA}}=({\omega_c+\omega_b})\pm\sqrt{({\omega_c-\omega_b})^2+4g^2}$
are the eigenfrequencies that were already obtained 
in Sec.~\ref{HM}. The main difference between the Lorentzians 
of Eq.~(\ref{spectrum10}) and Eq.~(\ref{ho2}) is that, in the 
latter, the corresponding $\Gamma_{\textsl{\tiny RWA}}$ of each 
$S_{x,y}^{\textsl{\tiny RWA}}(\omega,\Gamma)$
is the same, while in the former each $\Gamma_{x,y}$ 
is different. Later, we will see that this difference 
is the root of the asymmetric shape of the vacuum 
Rabi splitting of a system that is in the USC regimen 
of the light-matter interaction.
One thing in common between all the Lorentzian functions in 
the equations mentioned above is that, as a function of $\omega$, 
each of them is centered around the corresponding eigenfrequency, 
i.e., $S_{x,y}(\omega,\Gamma)$ is centered at $\omega_{x,y}$ 
and $S_{x,y}^{\textsl{\tiny RWA}}(\omega,\Gamma)$ 
at $\omega_{x,y}^{\textsl{\tiny RWA}}$.
This is because the spectrum's maximum intensity is expected
to happens at the resonant frequencies that can be obtained from 
the transitions between two energy levels of the coupled system; for example, by defining the polariton dispersion as
$\Delta E_{pq}^{mn}\equiv E_{mn}-E_{pq}=(m-p)\omega_x+(n-q)\omega_y$,
which is the energy associated with such transitions, 
$\Delta E^{10}_{00}=\omega_x$ and $\Delta E^{01}_{00}=\omega_y$
represent the transitions from the first two excited
states to the ground state of the Hopfield Hamiltonian.
\begin{figure}[t]
\begin{center}
\begin{large}
\includegraphics[width=15cm, height=10cm]{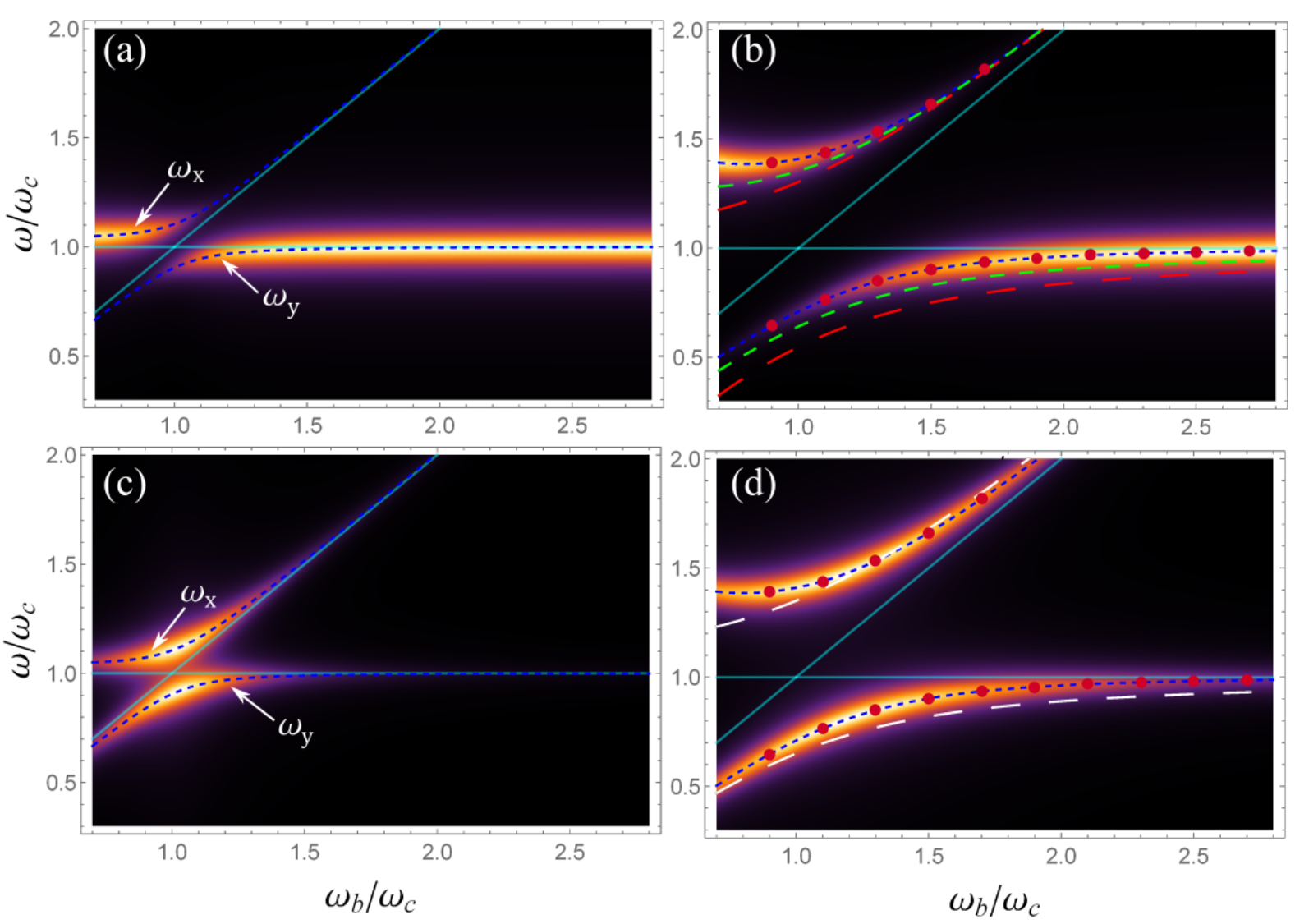}
\end{large}
\caption{Field spectra of the Hopfield model in the USC regime and within the single excitation manifold. For $\bf{(a)}$-$\bf{(b)}$ the initial state of the light-matter coupled system is $|1,0\rangle$, i.e.,  the excitation starts in the field while the matter is in its ground state; for $\bf{(c)}$-$\bf{(d)}$ we have the opposite initial state, $|0,1\rangle$. Density plots of $\bf{(a)}$,$\bf{(c)}$ are for $g=0.1\omega_c$ and $\bf{(b)}$,$\bf{(d)}$ for $g=0.35\omega_c$, in all cases the diamagnetic constant is $D=g^2/\omega_b$ and $\Gamma=0.05\omega_c$. When the normalized coupling constant $g/\omega_c$ increases, the separation of the avoiding crossing between the polariton branches also increases. Brighter (darker) colors are the regions where the intensity of the spectrum is high (low). Cyan solid lines are the bare frequencies $\omega_{c}$ and $\omega_b$ of the uncoupled ($g=0$) system. All dashed lines are the polaritonic dispersions representing the energy transition between the first two excited states of $\hat{H}_{\rm Hopfield}$ with the ground state. For the blue dashed lines, we set $D=g^2/\omega_b$ and the polariton dispersions coincide with the values where the spectrum's intensity is maximum, depicted as the red dots in $\bf{(b)}$,$\bf{(d)}$. For the red dashed lines, we set $D=0$, corresponding to the case of Fig.\ref{tutti}$\bf{(c)}$. The green dashed lines of $\bf{(b)}$ are for $D=0.5g^2/\omega_b$, while in $\bf{(d)}$ the white dashed lines are for $D=g_2=0$ and $g_1=g$, i.e., it is the situation where only the corotating terms are present in the Hopfield Hamiltonian. See the main text for more details.}
\label{fig3}
\end{center}
\end{figure}

In Fig.~\ref{fig3}$(a)$-$(b)$, we show density-plots of 
the spectrum of the Hopfield model,
$S(\omega,\Gamma)=S_x(\omega,\Gamma)+S_y(\omega,\Gamma)$,
as a function of $\omega_b/\omega_c$ with a normalized coupling strength $g/\omega_c$ equal to $0.1$ (a) 
and $0.35$  (b), i.e., the system is within the USC regime. 
Brighter (darker) colors represent the regions where the intensity of 
the spectrum is high (low). Near resonance ($\omega_b/\omega_c=1$),
one can see the characteristic avoiding crossing between the upper and
lower polaritons branches of the coupled system.
As expected, when the normalized coupling $g/\omega_c$ increases, 
the separation between these two branches also increases. 
In Fig.~\ref{fig3}$(a)$-$(b)$ we also show, as dashed-lines, the 
corresponding polariton dispersions $\Delta E_{00}^{10}$ and 
$\Delta E_{00}^{01}$ 
as a function of $\omega_b/\omega_c$ and for different values of the 
diamagnetic constant $D$. For example, by considering the TRK sum 
rule~\cite{Wang1999,Nataf2010}, we have that 
$D=g^2/\omega_b$~\cite{Ciuti2005,FriskKockum2019} (blue dashed-lines).
The blue-dashed lines fit very well within the regions where 
the spectrum's intensity is maximum, see the red dots.
The green dashed lines are for $D=d g^2/\omega_b$, where 
$d\in(0,1)$ is a prefactor that effectively reduces the diamagnetic 
term's strength. Such a value of $D$ represents a modification 
of the standard Hopfield model under the TRK sum rule; it was introduced very recently in~\cite{Keller2020} to explain the experimental data of Landau polaritons in highly nonparabolic two-dimensional gases in the USC regime; 
for Fig.~\ref{fig3}$(b)$ we set $d=0.5$. The red dashed lines 
are for $D=0$ and represent the case described in Fig.~\ref{tutti}$(c)$, 
i.e., a situation where the corotating and counterrotating terms 
are taken into account, but the diamagnetic term is not present.
The cyan solid lines represent the bare frequencies $\omega_{c}$
and $\omega_b$ of the system; in particular, the horizontal 
cyan line corresponds to the field frequency $\omega_c$.
It is easy to show that in the dispersive regime 
($\omega_b/\omega_c\gg 1$), $\omega_{x,y}\approx\omega_{b,c}$;
that is why for an initial state like $|1,0\rangle$ the spectrum 
of the field is brighter near the horizontal cyan line representing 
$\omega_c$. In the dispersive regime there is little exchange of 
energy between the field and the matter part.
Therefore, any initial excitation in the field will remain mainly 
there and should be captured by the EW spectrum around $\omega_c$.

As a complementary case, we now consider that in the 
Hopfield model the field starts in the vacuum and the 
matter part in its first excited state, i.e., the 
entire initial state is
$|0,1\rangle\equiv|0\rangle\otimes|1\rangle$, 
opposite situation of the previous example.
In this case, the field's autocorrelation function 
will be (see \ref{app2}):
$\langle 0,1|\hat{a}^{\dagger }(t_1)\hat{a} (t_2)|0,1\rangle= f^*_2( t_1 )  f_2( t_2 )+f^*_3( t_1 )  f_3( t_2 )+2f^*_4( t_1 )  f_4( t_2 )$, where, as in the previous example, we have
assumed $g_1=g_2=g$.
In the long-time limit, the corresponding EW spectrum can be written as
$S'(\omega,\Gamma)\equiv {S}'_x(\omega,\Gamma)+{S}'_y(\omega,\Gamma)$,
where
\begin{align}\label{spectrum01}
    {S}'_{x,y}(\omega,\Gamma) &=\frac{\Gamma'_{x,y}}{\Gamma ^2+(\omega-\omega _{x,y})^2},
    \,\,\,\,\Gamma'_{x,y} =\frac{\Gamma  h_{xy} }{32 \omega _b \omega _c \omega _{x,y}^2}(3\omega_b^2-2\omega_b\omega_{x,y}+3\omega_{x,y}^2) (\omega _c+\omega _{x,y})^2,
\end{align}
$h_{xy}=(4 \lambda ^2)(4 \lambda ^2+\Omega ^2)^{-1}$, $\Omega ={\omega_c}^2+4 D\omega_c-{\omega_b}^2$ 
and $\lambda =2 g \sqrt{\omega_b \omega_c}$.
Once again, we obtain that for the single 
excitation regime, but with the initial state $|0,1\rangle$, the 
Hopfield model's spectrum can be well approximated by the 
sum of two Lorentzian $S'_{x,y}(\omega,\Gamma)$ having different
$\Gamma'_{x,y}$.

For the initial state $|0,1\rangle$, Figs.~\ref{fig3}$(c)$-$(d)$ 
show density-plots of the spectrum of the field 
$S'(\omega,\Gamma)=S_x'(\omega,\Gamma)+S_y'(\omega,\Gamma)$.
We set the normalized coupling $g/\omega_c$ as $0.1$ (c) and 
$0.35$ (d). Like in the previous case, dashed-lines are the 
polaritonic dispersion from the first two excited states to 
the ground state. In particular, blue dashed-lines are for $D=g^2/\omega_b$ while 
white dashed-lines are for $D=g_2=0$ and $g_1=g$, i.e., the 
white dashed-lines represent the situation described in 
Fig.~\ref{tutti}$(d)$ where only corotating terms in the 
Hamiltonian are considered. As expected, the blue dashed lines 
match the values where the intensity of the spectrum is 
maximum and illustrated by the red dots in $(d)$. 
Notice that, contrary to Fig.~\ref{fig3}$(a)$-$(b)$, the 
intensity of the spectrum is now higher around the avoiding 
crossing region (near resonance). Moreover, in the dispersive 
limit, the spectrum's intensity vanishes nearby both solid 
cyan lines. Such behavior occurs because the initial state is
$|0,1\rangle$ and, therefore, there will not be enough energy 
to be captured by the spectrum of the field when $\omega_b/\omega_c\gg 1$.
In contrast, near resonance, there is a strong exchange of energy
between light and matter; thus, high values in the intensity 
of the field's spectrum can be obtained.

\subsection{Asymmetries in the vacuum Rabi splitting}

\begin{figure}[t]
\centering
{\includegraphics[width=7.5cm, height=5cm]{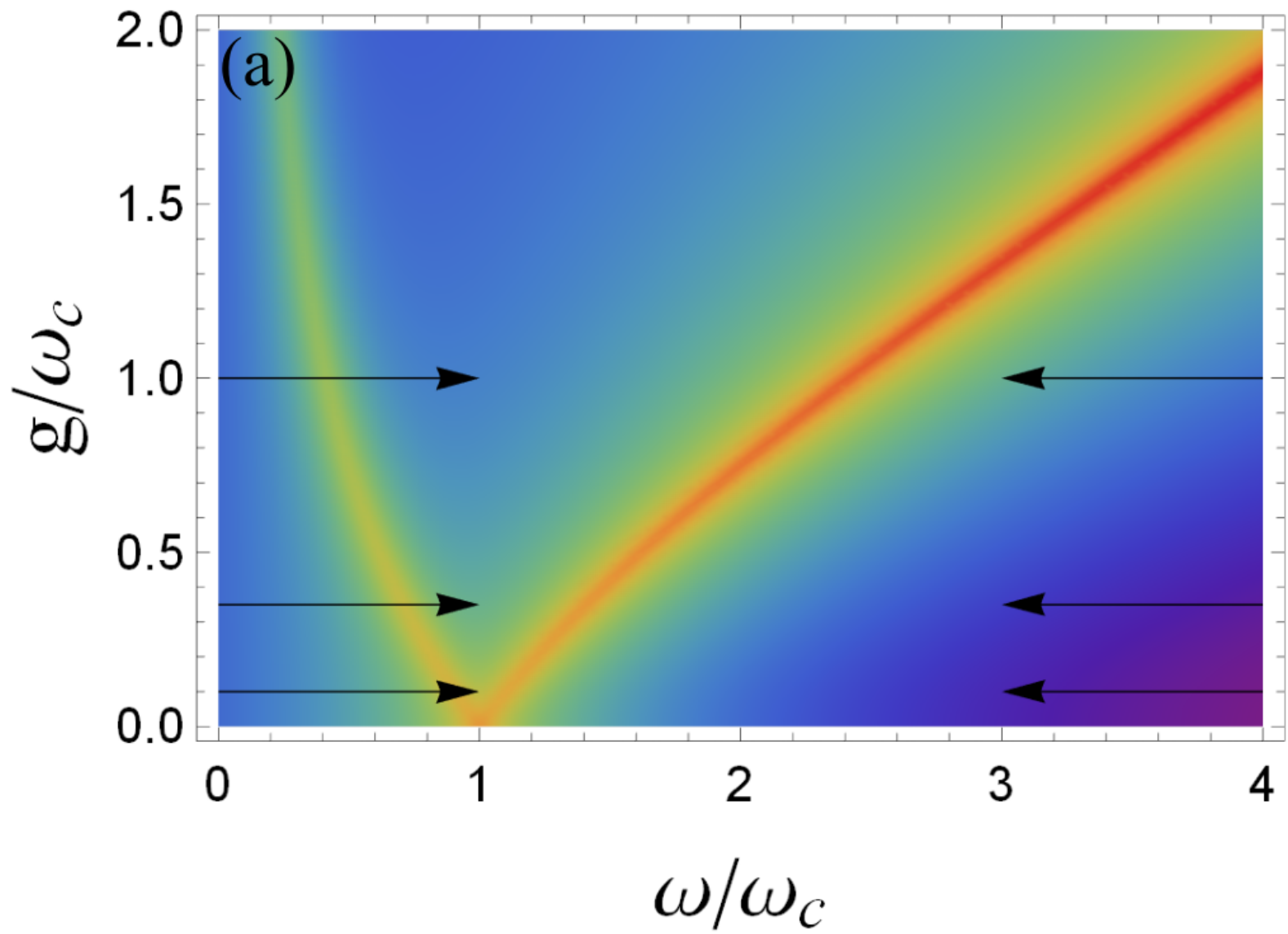}}
{\includegraphics[width=7.5cm, height=4.9cm]{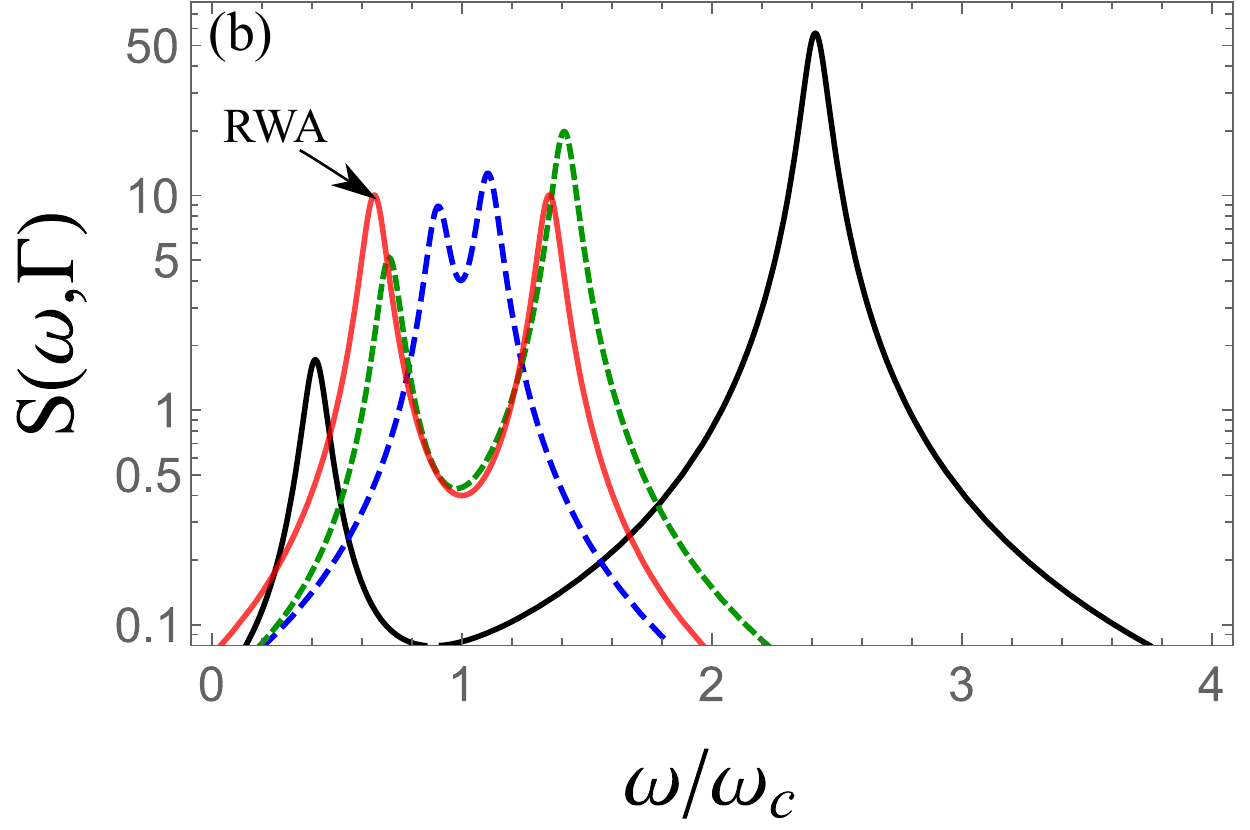}}
\caption{
Vacuum Rabi splitting (VRS) of the Hopfield model. $\bf{(a)}$ Shows the spectrum of the field $S(\omega,\Gamma)=S_x(\omega,\Gamma)+S_y(\omega,\Gamma)$, cf. Eq.(\ref{spectrum10}), as a function of the normalized coupling strength $g/\omega_c$ and frequency $\omega/\omega_c$, at resonance conditions, $\omega_c=\omega_b$. Red (blue) colors are the regions where the intensity of the spectrum is high (low).  In the USC regime the VRS is no longer symmetric with respect to the resonance frequency and the heights of each polariton branch. In fact, for large coupling strengths, the diamagnetic term in $\hat{H}_{\rm Hopfield}$ is dominant, and the VRS tends to a single polariton branch, which can be interpreted as an effective decoupling effect~\cite{DeLiberato2014}. $\bf{(b)}$ Shows, in a semi-log scale, the VRS for three different values of $g/\omega_c$ indicated by the dark arrows of $\bf{(a)}$. Blue dashed line is for $g=0.1\omega_c$, green dashed line for $g=0.35\omega_c$ and black solid line for $g=\omega_c$. In particular, the red solid line is for $g=0.35\omega_c$, but when the rotating-wave approximation (RWA) has been applied to $\hat{H}_{\rm Hopfield}$. Under this approximation the VRS is fully symmetric. \textcolor{black}{In all cases the diamagnetic constant is $D=g^2/\omega_b$ and $\Gamma=0.05\omega_c$.}
}
\label{fig4}
\end{figure}

From previous results, we can quickly obtain the 
vacuum Rabi splitting (VRS) of the light-matter interaction
in the Hopfield model when the bare frequencies of the field
and matter, $\omega_{c}$ and $\omega_{b}$, are resonant, 
$\omega_c=\omega_b$. In particular, in Fig.~\ref{fig4}$(a)$ 
we show a density plot for the EW spectrum of the field 
$S(\omega,\Gamma)=S_x(\omega,\Gamma)+S_y(\omega,\Gamma)$,
as a function of the normalized coupling $g/\omega_c$
and frequency $\omega/\omega_c$, at resonance conditions
$\omega_c=\omega_b$ and with the diamagnetic constant
under the TRK sum rule $D=g^2/\omega_b$. 
Red (blue) colors are the regions where the spectrum's 
intensity is high (low). For $g> 0.1\omega_c$, we see 
that the spectrum is not symmetric with respect to the resonance 
frequency and heights because of $\Gamma_x\neq\Gamma_y$. 
This asymmetry and the separation between the peaks of the two 
Lorentzian, $S_{x}(\omega,\Gamma)$ and $S_{x}(\omega,\Gamma)$, 
increases for larger values of $g/\omega_c$
and it is more evident in Fig.~\ref{fig4}$(b)$; there we
show, in semi-log scale, the VRS for three different values 
of the normalized coupling constant indicated by the dark 
arrows of Fig.~\ref{fig4}$(a)$. 
In Fig.~\ref{fig4}$(b)$ the right
Lorentzian $S_x(\omega,\Gamma)$ becomes dominant while the 
left one $S_y(\omega,\Gamma)$ decreases as the coupling 
constant increases; the blue dashed line is for
$g=0.1\omega_c$, green dashed-line is for $g=0.35\omega_c$,
and the solid black line is for $g/\omega_c=1$. The first two 
cases are in the USC regime, and the last one starts to 
enter the DSC regime. 
For the sake of comparison, the red solid-line of 
Fig.~\ref{fig4}$(b)$ shows the VRS, but when the RWA 
has been applied, i.e., it represents 
$S_{\textsl{\tiny RWA}}(\omega,\Gamma)= S_x^{\textsl{\tiny RWA}}(\omega,\Gamma)+S_y^{\textsl{\tiny RWA}}(\omega,\Gamma)$
for $g=0.35\omega_c$; 
thus, under the RWA, the VRS will be symmetric to the heights 
and the resonance frequency.
We might understand the asymmetric behavior of the VRS by 
looking for the eigenfrequencies 
$\omega_{x,y}$ in the limit where the normalized coupling 
is large. For instance, using Eq.~(\ref{omega_dia}) 
it is easy to show that if $D=g^2/\omega_b$ and $g\gg 1$, then 
$\omega_x^2\approx 4g^2\omega_c/\omega_b$, $\omega_y^2\approx 0$ and
the eigenvalues $E_{mn}$ of $\hat{H}_{\rm Hopfield}$ 
simplify to those of a single harmonic oscillator
$E_{mn}\approx \tilde\omega_x(m+1/2)$, with 
$\tilde\omega_x\equiv 2g\sqrt{\omega_c/\omega_b}$. In such limit,
the EW spectrum of the field reduces to just one Lorentzian
give by $S(\omega,\Gamma)\approx \tilde{S}_x(\omega,\Gamma)$, where
\begin{equation}\label{aproximacion}
 \tilde{S}_x(\omega,\Gamma )\equiv   \frac{3\Gamma g^2/(2\omega_b\omega _c)}{\Gamma ^2+ (\omega -\tilde\omega_x)^2}.
\end{equation}	
Thus, we see that in the DSC regime of the Hopfield model one of the two characteristic peaks of the VRS tends to vanish 
and the other remains and get more prominent, actually, 
in Refs.~\cite{Maissen_2017,Bayer2017,Todorov2019} significant 
asymmetries of the VRS were experimentally obtained, 
confirming the breakdown of the RWA and the relevance of the
diamagnetic term in the USC and DSC regime.
Having in mind that a measurement of the separation between
these two peaks gives the information about how strong the
light-matter coupling is, one can interpret the above tendency 
into a single peak as an effective decoupling between light and
matter when the coupling constant increases, a counterintuitive 
effect predicted in Ref.~\cite{DeLiberato2014} and measured 
\textcolor{black}{in Ref.~\cite{Mueller2020} by reaching the 
astonishing record $g/\omega_c=1.8$ at room temperature}. 
The effective decoupling happens because the diamagnetic term,
which is proportional to $g^2$ under the TRK sum rule, is the dominant term in 
the DSC regime, and it can act as a potential barrier~\cite{FriskKockum2019} for the photonic field producing 
the collapse of the Purcell effect~\cite{DeLiberato2014}. 

It is essential to mention that the asymmetry of the VRS 
remains, but it is less noticeable (plots not shown) if one
uses $|0,1\rangle$ as the initial state in the calculations of the 
EW spectrum, 
i.e., the field starts in the vacuum state while matter contains 
one excitation; the corresponding spectrum is 
$S'(\omega,\Gamma)=S'_x(\omega,\Gamma)+S'_y(\omega,\Gamma)$.
This observation might explain why in a recent experimental 
demonstration of the vacuum Bloch-Siegert shift, the VRS looks 
almost symmetric even when $g/\omega_c=0.36$~\cite{Li2018NatPhot}.
In fact, in~\cite{Li2018NatPhot}, the USC is 
between a two-dimensional electron gas with the 
counter-rotating component of a terahertz photonic-crystal 
cavity's {\em vacuum} fluctuation field, realizing a strong-field 
phenomenon without a strong field~\cite{Bamba2019}.
Finally, we notice that similar spectral modifications have been 
pointed out in other theoretical 
works~\cite{Ciuti2006,Ridolfo_PRL,Blais2011,Cao_2011};
however, they use more involved dissipative models to 
obtain these results. Here, we remark that our analytical expressions allow us
to characterize in simple terms the spectral response of 
these two ultrastrongly coupled quantum systems.

\section{Quantum thermometry with the Hopfield model}\label{QT}
Similar to standard quantum thermometry studies using individual 
quantum probes as thermometers~\cite{Deffner2019book,Correa2015,QST_Steve}, 
we assume that our thermometer (made of the two interacting quantum systems 
described by $\hat H_{\rm Hopfield}$) is fully thermalized by allowing it to 
equilibrate with the sample to be probed.  The temperature $T$ of the sample is inferred from the thermal 
state of the probe $\hat\rho_T\equiv\exp(-\beta\hat H_{\rm Hopfield})/Z$, where
$Z$ is the partition function, $\beta=(k_BT)^{-1}$ is the inverse
temperature and $k_B$ is the Boltzmann constant. To determine the ultimate limit on the estimation of temperature, 
we need to resort to the quantum Cram\'er-Rao bound~\cite{Correa_Review} given by 
$\Delta T\geq 1/\sqrt{N\mathcal{F}(\hat\rho_T)}$, where $\Delta T$ 
is the uncertainty in the temperature, $N$ the number of independent 
measurements and $\mathcal{F}(\hat{\rho}_T)$ is the quantum Fisher 
information (QFI). 
For thermal states, the QFI can be written in terms 
of the variance of the thermometer's Hamiltonian 
$\mathcal{F}(\hat\rho_T)=(\Delta\hat{H}_{\rm Hopfield})^2/(k_B^2T^4)$,
where $(\Delta\hat{H}_{\rm Hopfield})^2\equiv\langle\hat{H}^2_{\rm Hopfield}\rangle-\langle\hat{H}_{\rm Hopfield}\rangle^2$.
We can rewrite the QFI in terms of the thermometer's 
heat capacity as $\mathcal{F}(\hat\rho_T)=C(T)/(k_B T^2)$, 
where $C(T)= (\Delta\hat{H}_{\rm Hopfield})^2/(k_B T^2)$,
which implies that for the single-shot scenario of 
$N=1$~\cite{Correa2015}, the signal-to-noise ratio, 
$T/\Delta T$, is upper bounded as $(T/\Delta T)^2\leq C(T)/k_B$.
Moreover, the heat capacity can also be obtained from the 
internal energy, $U$, as follow: $C(T)=\partial_T U$, with 
$U=k_BT^2\partial_T\ln Z$~\cite{Grainer1995}. 
Thus, we only need to know how the thermometer's partition 
function is to obtain the corresponding ultimate  
bound on the thermal sensitivity through the QFI. 
To compute $Z={\rm tr}\{\exp(-\beta\hat H_{\rm Hopfield})\}$, 
we will take advantage of the analytical solution of the 
Hopfield model given in Sec.~\ref{HM}. For instance, the trace
operation is independent of the representation, so we can 
calculate $Z$ in the normal mode representation where 
$\hat{H}_{\rm Hopfield}$ is diagonal, and it represents two 
decoupled oscillators with eigenfrequencies $\omega_{x,y}$,
Eq.~(\ref{polfre}). On such basis, the partition function of 
two independent quantum system factorizes as $Z=Z_xZ_y$, 
where $Z_{x,y}=\frac{1}{2}{\rm csch}(\beta\omega_{x,y}/2)$ 
is the partition function of a quantum harmonic oscillator
in a thermal state~\cite{Grainer1995}. We can efficiently compute 
the internal energy and the heat capacity as $U=U_x+U_y$ and 
$C(T)=C_x(T)+C_y(T)$, where 
$U_{x,y}=\frac{1}{2}\omega_{x,y}\coth(\beta\omega_{x,y}/2)$ and 
$C_{x,y}(T)=\frac{1}{4}k_B(\beta\omega_{x,y})^2{\rm csch}^2(\beta\omega_{x,y}/2)$.
Therefore, the QFI of the anisotropic Hopfield model acting as a
thermometer reduces to
$\mathcal{F}(\hat{\rho}_T)=\mathcal{F}_x(\hat{\rho}_T)+\mathcal{F}_y(\hat{\rho}_T)$,
where
\begin{equation}\label{fisherH}
\mathcal{F}_{x,y}(\hat{\rho}_T)=
\Big(\frac{\omega_{x,y}}{2k_BT^2}\Big)^2{\rm csch}^2\bigg(\frac{\omega_{x,y}}{2k_BT}\bigg)
\end{equation}
is the well known QFI of a harmonic 
oscillator~\cite{Correa2015,roman2019spectral}.

\begin{figure}[t]
\centering
\includegraphics[width=15cm, height=5.3cm]{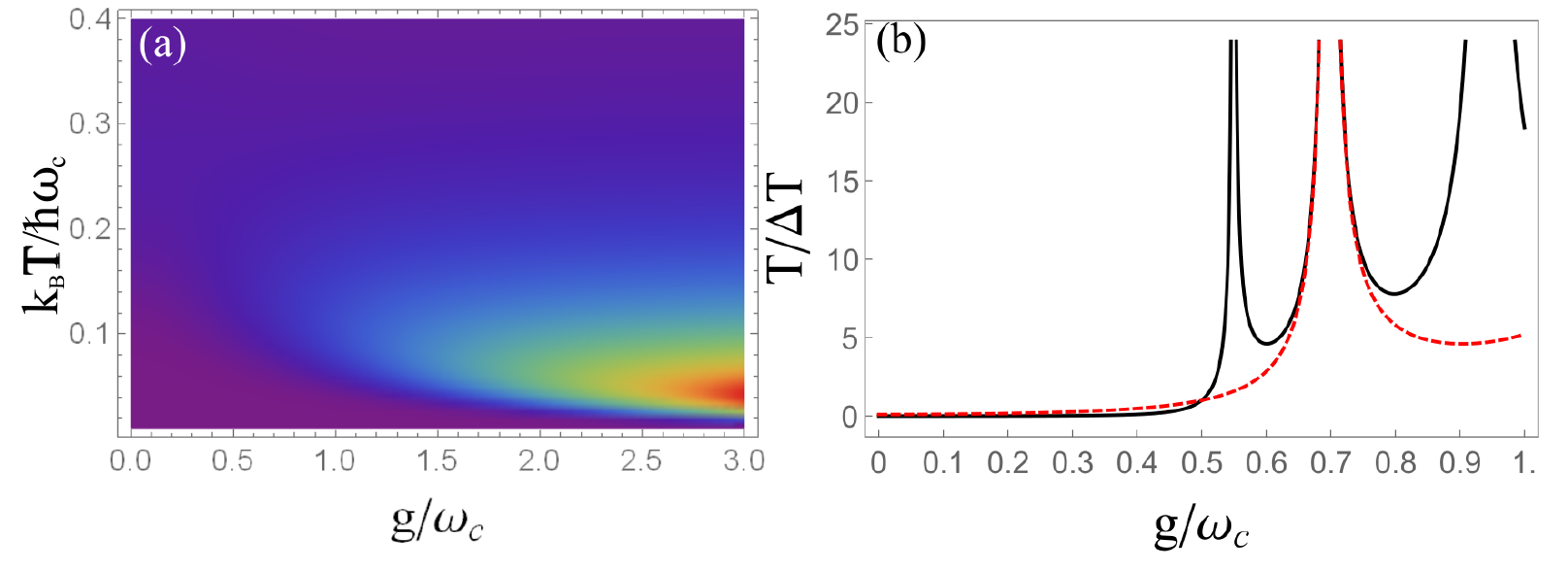}
\caption{
$\bf{(a)}$ Quantum Fisher information (QFI) of the composite quantum probe, $\mathcal{F}(\hat\rho_T)=\mathcal{F}_x(\hat\rho_T)+\mathcal{F}_x(\hat\rho_T)$, cf. Eq.~(\ref{fisherH}), as a function of the normalized coupling $g/\omega_c$ and the  \textcolor{black}{scaled temperature $k_BT/\hbar\omega_c$}. Red (blue) colors represent high (low) values of $\mathcal{F}(\hat\rho_T)$. The QFI substantially increases if the probe is in the USC regime. The set of parameters are those used in Fig.~\ref{tutti}({\bf b}), i.e., the system is on resonance $\omega_c=\omega_b$, $g_1=g_2=g$ and $D=g^2/\omega_b$. $\bf{(b)}$.  Signal-to-noise ratio, $T/\Delta T\equiv T\sqrt{\mathcal{F}(\hat\rho_T)}$, versus $g/\omega_c$, with a \textcolor{black}{scaled temperature $k_BT/\hbar\omega_c=0.05$} (black solid line) and \textcolor{black}{$k_BT/\hbar\omega_c=0.1$} (red dashed line). For this case $D=0$ and $\mathcal{F}(\hat\rho_T)$ is given by Eq.~(\ref{fisherD0}). At low temperatures a divergence of $T/\Delta T$ appears if the system is close to the critical point ($g_{\rm crit}\equiv \omega_c/2$) of the quantum phase transition described in Fig.~\ref{tutti}({\bf c}). If $g>g_{\rm crit}$, periodic divergences occur because one of hyperbolic functions of Eq.~(\ref{fisherD0}) changes to its trigonometric version. In principle, one could obtain arbitrarily high thermometry precision by exploiting the criticality of this quantum sensor.
}
\label{qfid0}
\end{figure}

In Fig.~\ref{qfid0}$(a)$, we show 
$\mathcal{F}_x(\hat{\rho}_T)+\mathcal{F}_y(\hat{\rho}_T)$
as a function of the temperature and the normalized coupling
using the same set of parameters of Fig.~\ref{tutti}$(b)$, 
i.e., a situation where the interacting systems are on resonance
$(\omega_c=\omega_b)$, the diamagnetic constant is obtained from 
the TRK sum rule ($D=g^2/\omega_b$), and the contribution of the 
co-rotating and counter-rotating terms is the same $(g_1=g_2=g)$; 
red (blue) colors represent high (low) values of the QFI. 
We can see that for any value of the temperature the 
QFI substantially increases, and the thermometry 
precision as well, when the thermometer operates in the 
USC and DSC regime. 
Similar results were obtained in Refs.~\cite{Campbell2017,Correa_Review} 
but for the particular case in which the RWA is 
used in the Hamiltonian of two coupled bosonic modes. 
If we replace $\omega_{x,y}$ by $\omega_{x,y}^{\textsl{\tiny RWA}}$ 
in Eq.~(\ref{fisherH}), we obtain the same expression of the QFI 
derived in~\cite{Campbell2017,Correa_Review}. 

Such enhancement in the estimation of the temperature is due to the fact that the 
gap between the corresponding energy levels of the thermometer 
decreases when the coupling constant increases, see 
Fig.~\ref{tutti}$(b)$, which makes it easy to populate more excited 
energy states. 
Therefore, in an extreme case where the gap vanishes, 
significant thermometry precision changes are to be 
expected~\cite{Correa_Review}. Interestingly, a vanishing gap 
occurs during the spectral collapse of Fig.~\ref{tutti}$(c)$, 
where $D=0$ and $\hat{H}_{\rm Hopfield}$ represents a 
Dicke-type Hamiltonian with a quantum phase transition at the 
critical point $g_{\rm crit}\equiv{\omega_c}/2$, see Sec.~\ref{HM}. 
In such a case, the eigenfrequencies of Eq.~(\ref{omega_dia}) 
simplify to $\omega^2_{x,y}=\omega_c^2\pm 2g\omega_c$, which
we substitute in Eq.~(\ref{fisherH}) to get
\begin{align}\label{fisherD0}
\mathcal{F}(\hat\rho_T)=
\frac{\omega_c^2+2 g\omega_c}{4k_B^2 T^4} {\rm csch}^2\bigg(\frac{\sqrt{\omega_c^2+2 g\omega_c}}{2 k_B T}\bigg)
+
\frac{\omega_c^2-2 g\omega_c}{4k_B^2 T^4}{\rm csch}^2\bigg(\frac{\sqrt{\omega_c^2-2 g\omega_c}}{2k_BT}\bigg). 
\end{align}
When $g=g_{\rm crit}$ the first term in the above equation vanishes
at low temperatures, while the second one is proportional to 
$T^{-2}$, making the QFI diverge. This divergence suggests that 
an arbitrarily high thermometry precision can be achieved in the
critical region's vicinity at low enough temperatures~\cite{Correa_Review}.
\textcolor{black}{However, since $C(T)=\mathcal{F}(\hat{\rho}_T)k_BT^2$, the $T^{-2}$ dependency of the QFI implies that $C(T)$ will remain constant even at $T=0$; this is, somehow, a contradiction with the Third Law of thermodynamics, where it is generally assumed that for any system with a finite energy gap, the heat capacity should vanish (exponentially) at $T=0$. In fact, this vanishing of the heat capacity is why it is so difficult to estimate low temperatures. To be more specific, the relative error $(\Delta T/T)^2\geq k_B/C(T)$ necessary diverges exponentially as $T\rightarrow 0$~\cite{Potts2019}. Nevertheless, finite quantum dissipation may restore the Third Law~\cite{Hanggi2005}, indicating that important effects resulting from the coupling to the environment may not be neglected.}
Remarkably, if $g>g_{\rm crit}$ several divergences in the QFI 
may appear, which is our second major result; such atypical behavior of the 
QFI when $g>g_{\rm crit}$ is due to the argument on the hyperbolic 
function in the second term of Eq.~(\ref{fisherD0}) which becomes complex \textcolor{black}{(purely imaginary)}, 
this changes the hyperbolic cosecant to its trigonometric version, which 
as a function of the normalized coupling has periodic divergences that
can be tailored depending on the temperature. Recall that in the region
$g>g_{\rm crit}$ the energy level structure of the coupled system is no 
longer discrete but continuous.
Using Eq.~(\ref{fisherD0}), we plot in Fig.~\ref{qfid0}$(b)$ the 
signal-to-noise ratio 
${T}/{\Delta T}\equiv T\sqrt{\mathcal{F}(\hat{\rho}_T)}$ as a function
of the normalized coupling; the solid black line shows the above-mentioned 
periodic divergences of the QFI at low temperatures, and these are modified
if the temperature increases, see the red dashed line.
\textcolor{black}{
We would like to point out that, at first glance, one would expect 
$T/\Delta T$ to diverge at the quantum critical point; however, 
this is not what we see in Fig.~\ref{qfid0}$(b)$. Actually, it is 
easy to show that each divergence in Fig.~\ref{qfid0}$(b)$ is located at 
${g}/\omega_c= \frac{1}{2}+2(k_BT n\pi)^2$, with $n\in \mathbb{N}$.
The physical reason is that at equilibrium, which is the case we 
are considering, the critical point is shifted by the temperature. 
Another well-known example where this fact occurs is the Dicke model, in which, by following a mean-force approach and an optimization procedure in the free energy of each two-level atom, one can obtain the critical coupling $\lambda_c$ of its equilibrium transition (see Eq.~(9) of \cite{Emanuele}): $\lambda_c=\frac{1}{2}\sqrt{\omega_c\omega_z\coth\left({\beta\omega_z}/{2}\right)}$, where $\omega_c$ and $\omega_z$ are the frequencies of the cavity field and the two-level atomic transition respectively and $\beta$ is the inverse temperature. This critical coupling reduces to $\lambda_c=\sqrt{\omega_c\omega_z}/2$ when $\beta\rightarrow\infty$, the expected result of the corresponding quantum phase transition.}
Similar results\textcolor{black}{, at zero temperature,} were recently 
obtained in~\cite{Garbe2020} but only for frequency-estimation protocols, 
where the criticallity of finite-component quantum optical probes was exploited.

\section{Conclusions}\label{Conc}
We have shown that the spectral response of two ultrastrongly 
coupled quantum systems (described by the Hopfield model) can be 
characterized by the Eberly-W\'odkiewicz (EW) physical spectrum. 
Through a non-trivial diagonalization procedure, we derived an exact 
analytical solution of the anisotropic Hopfield Hamiltonian, and we 
found the corresponding polaritonic frequencies, see Eq.~(\ref{polfre}), 
as well as the autocorrelation function of the field. 
The eigenfrequencies were useful to obtain a better description (compared 
with previous numerical results~\cite{FriskKockum2019}) of the energy level 
structure of the coupled system, including its quantum phase transition.
Simultaneously, the autocorrelation function was necessary to perform the 
two time-integrals of the EW spectrum. 
In the long-time limit, the resulting EW spectrum is time-independent
and it can be reduced to a simple expression, Eq.~(\ref{spectrum10}), 
that remains valid in the USC and DSC regime. 
We confirm that in the DSC regime, the vacuum Rabi 
splitting manifests large asymmetries that can be considered spectral 
signatures of the  counterintuitive decoupling effect.	

We also show that when the Hopfield model is used as a quantum 
thermometer, the corresponding ultimate bounds on the estimation of 
temperature are valid in the USC and DSC regime; 
regimes that, to the best of our knowledge, had not been explored 
yet in the field of quantum thermometry. 
For both regimes, the quantum Fisher information 
(QFI) of the Hopfield model substantially increases. 
Remarkably, when the coupled system (described by a Dicke-type Hamiltonian) 
performs a superradiant phase transition (SPT),
the QFI displays periodic divergences that allow it to have, in principle, 
arbitrary high thermometry precision. 
This thermometry result depends on the fact that the diamagnetic 
term should not be present in the USC and DSC regime; 
therefore, it might not be possible to test it 
in quantum optics experiments where light interacts with natural 
atoms. However, it could be implemented in condensed 
matter systems using matter-matter interactions;
for instance, Dicke cooperativity was experimentally realized 
in~\cite{Li2018} with spin-magnon interactions in the USC regime.
Moreover, recent terahertz magnetospectroscopy 
experiments~\cite{Makihara_arXiv_20,Makihara20,Peraca20} suggest 
that the magnonic version of the SPT is within reach~\cite{Bamba_2020}.

Using the Hopfield 
model's analytical solution presented in this work, 
it would be possible to derive the corresponding 
microscopic (global) Lindblad master 
equation~\cite{Hofer_2017}. 
With the master equation at hand, one could investigate 
the impact of the USC on the energy transfer dynamics 
or the thermodynamics properties~\cite{Rabl_USC}.
%
Moreover, it is known that the output power of a particular 
class~\cite{Latune2020} of quantum heat engines is determined 
by the heat capacity of their working substance~\cite{Abiuso2020};
therefore, if we use the Hopfield model as a working medium, 
significant changes in its heat capacity are expected to enhance 
the maximal output power of such a critical heat engine operating 
in the USC regime~\cite{Campisi2016}. These are a few exciting 
topics of research that we will address in future work.

\textcolor{black}{{\it Note added}. After completing this manuscript, we became aware of related work~\cite{Ivette2019}, which considers a general solution of two interacting harmonic oscillators resembling the anisotropic Hopfield model.}

\section*{Acknowledgments}
M.S.-M. would like to express her gratitude to CONACyT, Mexico for her Scholarship.

\appendix

\section{Diagonalization of the anisotropic Hopfield Hamiltonian}\label{app1}
In this Appendix, we describe all the necessary steps to diagonalize the anisotropic ($g_1\neq g_2$) Hopfield Hamiltonian $\hat{H}_{\rm Hopfield}$ given in Eq.~(\ref{hopfield}) of the main text. First, we make a change of sign on the third and four term of $\hat{H}_{\rm Hopfield}$ by means of the unitary transformation $\hat{T}=\exp( -i{\pi}\hat{b}^\dagger \hat{b}/2)$, i.e., $\hat{T} \hat{H}_{\rm Hopfield}\hat{T}^{\dagger}\equiv\hat{H}_1$, where $\hat T\hat b\hat T^\dagger=i\hat b$, $\hat T\hat b^\dagger\hat T^\dagger=-i\hat b^\dagger$ such that
\begin{equation}\label{A1.0020}
\hat{H}_1=\omega_c \hat{a}^\dagger \hat{a} + \omega_b \hat{b}^\dagger \hat{b}
+ g_1 (\hat{a}\hat{b}^\dagger + \hat{a}^\dagger \hat{b} ) 
+ g_2 (\hat{a}^\dagger\hat{b}^\dagger + \hat{a} \hat{b} ) 
+ D ( \hat{a}+\hat{a}^\dagger)^2 .
\end{equation}
Next, we rewrite $\hat H_1$ in terms of the hermitian operators $\hat{x}$, $\hat{p}_x$, $\hat{y}$ and $\hat{p}_y$, defined by the relations
$\hat{a}=\sqrt{{\omega_c}/{2}} \big(\hat{x}+{i}\hat{p}_x/\omega_c\big)$, $\hat{b}=\sqrt{{\omega_b}/{2}} \big(\hat{y}+{i} \hat{p}_y/{\omega_b}\big)$ and their adjoint operators, evidently $[\hat{x},\hat{p}_x]=i$ and $[\hat{y},\hat{p}_y]=i$. This yields
\begin{equation}\label{A1.0050}
	\hat{H}_1=\left( \hat{p}_x^2 + \hat{p}_y^2
	+\Omega_x  \hat{x}^2 + \Omega_y^2  \hat{y}^2\right)/2 
	+\lambda_1 \hat{x}\hat{y}  +\lambda_2 \hat{p}_x \hat{p}_y,
\end{equation}
where we have defined the quantities
$\Omega_x^2$=$\omega_c^2+4 D \omega_c$, $\Omega_y^2$=$\omega_b^2$, 
$\lambda_{1,2}=\left( g_1 \pm g_2 \right)({\omega_c \omega_b})^{\pm {1}/{2}}$.
Now, using the unitary transformation $\hat{R}=\exp\big[ i {\pi}( \hat{x} \hat{p}_y - \hat{y} \hat{p}_x )/4 \big]$, we perform a $\pi/4$ rotation in~(\ref{A1.0050}), i.e., $\hat{R} \hat{H}_1 \hat{R}^\dagger\equiv\hat{H}_2$, which removes the coupling term $\lambda_2\hat{p}_x\hat{p}_y$ of~(\ref{A1.0050}):
\begin{equation}\label{A1.0080}
	\hat{H}_2=
	\left[ \left(1+\lambda_2 \right)\hat{p}_x^2 + \left(1-\lambda_2 \right)\hat{p}_y^2 + \Omega_1^2 \hat{x}^2 + \Omega_2^2 \hat{y}^2
	\right]/2 + \tilde{\lambda} \hat{x} \hat{y},
\end{equation}
where
$\Omega_1^2 =\big({\Omega_x^2+\Omega_y^2}\big)/{2}+\lambda_1$, 
$\Omega_2^2=\big({\Omega_x^2+\Omega_y^2}\big)/{2} - \lambda_1$ and
$\tilde{\lambda}=\big({\Omega_x^2-\Omega_y^2}\big)/{2}$.
Note that, in order to obtain~(\ref{A1.0080}), we have used the follow identities	
\begin{eqnarray}
\hat R\hat p_x\hat R^\dagger=(\hat p_x-\hat p_y)/2,\quad
\hat R\hat x\hat R^\dagger=(\hat x-\hat y)/2,\\
\hat R\hat p_y\hat R^\dagger=(\hat p_x+\hat p_y)/2,\quad
\hat R\hat y\hat R^\dagger=(\hat x+\hat y)/2.\label{RyRdagg}
\end{eqnarray}
By using $\hat{S}=\exp\big[i {r_1} \left( \hat{x} \hat{p}_x + \hat{p}_x \hat{x}\right)/2  \big]   \exp\big[i {r_2}\left( \hat{y} \hat{p}_y + \hat{p}_y \hat{y}\right)/2\big]$,
where $r_1=\ln\sqrt{1+\lambda_2}$ and $r_2=\ln\sqrt{1-\lambda_2}$,
we carry out a squeezing transformation $\hat{S}\hat{H}_2\hat{S}^\dagger\equiv\hat{H}_3$.
The actions of this squeezing transformation,
\begin{eqnarray}
\hat{S}\hat{x}\hat{S}^\dagger=\hat{x}(1+\lambda_2),\quad \hat{S}\hat{p}_x\hat{S}^\dagger=\hat{p}_x(1+\lambda_2)^{-1},\\\hat{S}\hat{y}\hat{S}^\dagger=\hat{y}(1-\lambda_2),\quad \hat{S}\hat{p}_y\hat{S}^\dagger=\hat{p}_y(1-\lambda_2)^{-1},
\end{eqnarray}
allow us to get
$\hat{H}_3=\left( \hat{p}_x^2 + \hat{p}_y^2 + w_1^2 \hat{x}^2 + w_2^2 \hat{y}^2	\right)/2 + \lambda \hat{x} \hat{y}$,
where the new definitions
$w_{1,2}^2=\left(1\pm \lambda_2 \right)(\Omega_x^2+\Omega_y^2\pm 2\lambda_1)/2$
and $\lambda=({1-\lambda_2^2})^{\frac{1}{2}}({\Omega_y^2-\Omega_x^2})/{2}$
have been used.
Above, we have omitted the discussion about the validity of this squeezing transformation.
It is clear that if $\lambda_2<-1$  or $\lambda_2>1$ the transformation can not be done
because one the parameters $r_1$ or $r_2$ become undetermined.
These conditions correspond, in $\hat{H}_2$, to have effective negative masses of the 
corresponding oscillators.

As last step, we  use $\hat{R}_\theta=\exp\big[ i \theta ( \hat{x} \hat{p}_y - \hat{y} \hat{p}_x ) \big]$
with $\tan (2\theta)={2\lambda}({w_1^2-w_2^2})^{-1}$ to make another rotation, $\hat{R}_\theta\hat H_3\hat{R}_\theta^\dagger\equiv \hat H_{\rm diag}$, 
and remove the coupling term $\lambda\hat{x}\hat{y}$ of $\hat{H}_3$. This yields
$\hat H_{\rm diag}=(\hat{p}_x^2+ \omega_x^2\hat{x}^2)/2 +(\hat{p}_y^2 + \omega_y^2 \hat{y}^2)/2$,
which is the Hamiltonian of two decoupled quantum harmonic oscillators 
with eigenvalues $E_{mn}=\omega_x(m+1/2)+\omega_y(n+1/2)$ and frequencies
$2\,\omega_{x,y}^2=({w_1^2+w_2^2})\pm(w_1 - w_2)^{-1}({ w_1^2-w_2^2+4\lambda^2 })^{{1}/{2}}$.
Here, $n$ and $m$ are non-negative integers.
To obtain such diagonal Hamiltonian, $\hat{H}_{\rm diag}$,
 we have used the follow identities:
\begin{eqnarray}
\hat{R}_\theta\hat{x}\hat{R}_\theta^\dagger=\hat{x}\cos\theta-\hat{y}\sin\theta,\quad
\hat{R}_\theta\hat{p}_x\hat{R}_\theta^\dagger=\hat{p}_x\cos\theta-\hat{p}_y\sin\theta,\\
\hat{R}_\theta\hat{y}\hat{R}_\theta^\dagger=\hat{x}\sin\theta+\hat{y}\cos\theta,\quad
\hat{R}_\theta\hat{p}_y\hat{R}_\theta^\dagger=\hat{p}_x\sin\theta+\hat{p}_y\cos\theta.
\end{eqnarray}
By substituting $w_{1,2}$ in $\omega_{x,y}$ we obtain
Eq.~(\ref{polfre}) of the main text.
Note that when $w_1=w_2$ the angle 
$\theta$ is not undefined, actually in such limit 
$\theta\rightarrow\pi/4$ and $\omega_{x,y}^2=w_1^2\pm\lambda$.
In summary, $\hat{H}_{\rm diag}=\hat{R}_\theta\hat{S}\hat{R}\hat{T}\hat{H}_{\rm Hopfield}\hat{T}^\dagger\hat{R}^\dagger\hat{S}^\dagger\hat{R}_\theta^\dagger$ and $E_{mn}$ are, in fact, the exact eigenvalues of the anisotropic Hopfield Hamiltonian.

\section{Temporal evolution of the field operators and the autocorrelation function} \label{app2}
Here, we compute the time evolution of the field 
operator $\hat{a}$ and the corresponding autocorrelation
function needed to calculate the physical spectrum 
in Eq.~\eqref{spectrum} of the main text. 
The field operator is written as
$\hat{a}\left( t \right) = \hat{U}\left( t \right)^\dagger \hat{a}   \hat{U}\left( t \right)$,
where $\hat{U}(t)$ is the time evolution operator
$\hat{U}\left( t \right)=\exp(- it \hat{H}_{\textrm{Hopfield}} )$
and we have denoted the value of 
$\hat{a}\left( t\right) $ at time zero, 
$\hat{a}\left( t=0 \right) $, just as  $\hat{a}$.
At the end of~\ref{app1} we find that 
$\hat{H}_{\rm Hopfield}=\hat{T}^\dagger\hat{R}^\dagger\hat{S}^\dagger\hat{R}^\dagger_\theta
\hat{H}_{\rm diag}\hat{R}_\theta\hat{S}\hat{R}\hat{T}$,
however, if $g_1=g_2=g$ then $\lambda_2=0$, $r_{1,2}=0$
and the squeezing transformation $\hat{S}$ becomes the 
identity operator, see the text bellow~(\ref{A1.0050}) 
and~(\ref{RyRdagg}).
In such situation $\hat{H}_{\rm Hopfield}$ reduces to
$\hat{H}_{\rm Hopfield}=\hat{T}^\dagger\hat{R}^\dagger\hat{R}^\dagger_\theta\hat{H}_{\rm diag}\hat{R}_\theta\hat{R}\hat{T}$.
If the two rotations $\hat{R}_\theta\hat{R}$ are written as 
one rotation $\hat{R}_\phi$, with the angle $\phi\equiv\theta+\pi/4$,
the Hamiltonian simplifies even further
$\hat{H}_{\rm Hopfield}=\hat{T}^\dagger\hat{R}^\dagger_\phi\hat{H}_{\rm diag}\hat{R}_\phi\hat{T}$.
The evolution operator 
can be written as a product of unitary transformations
$\hat{U}(t)=\hat{T}^\dagger\hat{R}^\dagger_\phi\exp(-it\hat{H}_{\rm diag})\hat{R}_\phi\hat{T}$.
Thus, the field operator reads as
$\hat{a}\left( t \right) =\hat{T}^\dagger\hat{R}_\phi^\dagger  e^{ it  \hat{{H}}_{\rm diag}  }   \hat{R}_\phi\hat{T}  \hat{a}  \hat{T}^\dagger\hat{R}_\phi^\dagger e^{ -it  \hat{{H}}_{\rm diag}  } \hat{R}_\phi\hat{T}$.
So we have to make one by one all of these transformations,
as an example, it is easy to see that the first one is 
$\hat{T}\hat{a}\hat{T}^\dagger=\hat{a}$,
however, the next unitary transformations are too cumbersome 
to be shown here. The result of making all  these 
transformations to the field operator is the following:
$\hat{a}( t ) = f_1( t )  \hat{a} + f_2( t ) \hat{a}^\dagger +f_3( t ) \hat{b} + f_4( t ) \hat{b}^\dagger$,
where the functions $f_j(t)$ are defined as 
$f_j(t)$=$\sum_{k=1}^2\exp(\pm i\omega_x t)\mu_{jk}+\sum_{k=3}^4\exp(\pm i\omega_y t)\mu_{jk}$
and the coefficients $\mu_{jk}$ are:
\begin{align}
\mu_{11}=&-\frac{\cos ^2(\phi ) (\omega _c-\omega _x){}^2}{4 \omega _c \omega _x}, \qquad \qquad \qquad 
\mu_{12}= \frac{\cos ^2(\phi ) \left(\omega _c+\omega _x\right){}^2}{4 \omega _c \omega _x}, \nonumber \\ 
\mu_{13}=&-\frac{\sin ^2(\phi ) \left(\omega _c-\omega _y\right){}^2}{4 \omega _c \omega _y}, \qquad \qquad \qquad \nonumber
\mu_{14}=\frac{\sin ^2(\phi ) \left(\omega _c+\omega _y\right){}^2}{4 \omega _c \omega _y}, \\ \label{mu}
\mu_{21}=& \frac{\cos ^2(\phi ) \left(\omega _c-\omega _x\right) \left(\omega _c+\omega _x\right)}{4 \omega _c \omega _x}, \qquad \qquad \nonumber
\mu_{22}=\frac{\cos ^2(\phi ) \left(\omega _x^2-\omega _c^2\right)}{4 \omega _c \omega _x}, \\ \nonumber
\mu_{23}=&   \frac{\sin ^2(\phi ) \left(\omega _c-\omega _y\right) \left(\omega _c+\omega _y\right)}{4 \omega _c \omega _y}, \qquad \qquad  
\mu_{24}= \frac{\sin ^2(\phi ) \left(\omega _y^2-\omega _c^2\right)}{4 \omega _c \omega _y}, \\ \nonumber
\mu_{31}=&  \frac{i \sin (2 \phi ) \left(\omega _b-\omega _x\right) \left(\omega _c-\omega _x\right)}{8 \omega _x \sqrt{\omega _b \omega _c}}, \qquad\qquad 
\mu_{32}=-\frac{i \sin (2 \phi ) \left(\omega _b+\omega _x\right) \left(\omega _c+\omega _x\right)}{8 \omega _x \sqrt{\omega _b \omega _c}}, \\ \nonumber 
\mu_{33}=&   \frac{i \sin (2 \phi ) \left(\omega _b-\omega _y\right) \left(\omega _y-\omega _c\right)}{8 \omega _y \sqrt{\omega _b \omega _c}} , \qquad  \qquad
\mu_{34}= \frac{i \sin (2 \phi ) \left(\omega _b+\omega _y\right) \left(\omega _c+\omega _y\right)}{8 \omega _y \sqrt{\omega _b \omega _c}}             , \\ \nonumber
\mu_{41}=&    \frac{i \sin (2 \phi ) \left(\omega _b+\omega _x\right) \left(\omega _c-\omega _x\right)}{8 \omega _x \sqrt{\omega _b \omega _c}}            , \qquad \qquad
\mu_{42}=  -\frac{i \sin (2 \phi ) \left(\omega _b-\omega _x\right) \left(\omega _c+\omega _x\right)}{8 \omega _x \sqrt{\omega _b \omega _c}}, \nonumber \\ 
\mu_{43}=&     -\frac{i \sin (2 \phi ) \left(\omega _b+\omega _y\right) \left(\omega _c-\omega _y\right)}{8 \omega _y \sqrt{\omega _b \omega _c}}           , \qquad \quad
\mu_{44}=  \frac{i \sin (2 \phi ) \left(\omega _b-\omega _y\right) \left(\omega _c+\omega _y\right)}{8 \omega _y \sqrt{\omega _b \omega _c}}.
\end{align}
The frequencies $\omega_{x,y}$ are those given in 
Eq.~(\ref{omega_dia}) of the main text and the
angle $\phi$ can be written as 
$\tan(2\phi)={2\lambda}({\omega_x^2-\omega_y^2+4D\omega_c})^{-1}$
with $\lambda=2g\sqrt{\omega_c\omega_b}$.
Evidently, $\hat{a}^\dagger \left(t\right)= f^*_1\left( t \right)  \hat{a}^\dagger + f^*_2\left( t \right) \hat{a} +f^*_3\left( t \right) \hat{b}^\dagger + f^*_4\left( t \right) \hat{b}$. 

Finally, the autocorrelation function for an initial product 
state $|n,m\rangle\equiv|n\rangle\otimes|m\rangle$, where
$n$~($m$) means that there are $n$~($m$) excitations in the 
field (matter) part satisfying $n\neq m$, is given by
$\langle n,m|\hat{a}^{\dagger }(t_1)\hat{a} (t_2)|n,m\rangle=\big[f^*_1( t_1 )  f_1( t_2 )+f^*_2( t_1 )  f_2( t_2 )\big]n
+\big[f^*_3( t_1 )  f_3( t_2 )+f^*_4( t_1 )  f_4( t_2 )\big]m+f^*_2( t_1 )  f_2( t_2 )+f^*_4( t_1 )  f_4( t_2 )$.
Thus, we use this result to compute the physical 
spectrum of the Hopfield model, given in 
Eq.~(\ref{spectrum}), for the two examples of 
initial states that are considered along Sec.~\ref{SHM}
of the main text.

\section*{References}

\bibliographystyle{ieeetr} 
\bibliography{Ref2}

\end{document}